\documentclass[twocolumn,10pt]{tsfp}
\usepackage{amsmath}
\usepackage{flushend}
\usepackage{graphicx}
\usepackage[authoryear,round]{natbib}
\usepackage{fancyhdr, subcaption}
\usepackage{xcolor}

\newcommand{\emaila}{arturo.arosemena@ntnu.no}
\newcommand{\emailb}{procacci@kth.se}
\newcommand{\emailc}{simone.digiorgio@cnr.it}
\newcommand{\emaild}{jannike.solsvik@ntnu.no}
\newcommand{\emaile}{msey@kth.se}

\title{Turbulence anisotropy in a bubbly vertical channel flow with topological changes}

\renewcommand{\expauthors}{1.1}

\author{Arturo A. Arosemena
    \affiliation{
	Department of Chemical Engineering\\
	Norwegian University of Science and Technology\\
	Sem Sælands vei 6, Trondheim, Trøndelag, 7491, Norway\\
    \emaila
    }	
}

\author{Davide Procacci
          \affiliation{
        FLOW, Department of Engineering Mechanics\\
	KTH Royal Institute of Technology\\
	Osquars backe 18, Stockholm, Stockholms län, 114 28, Sweden\\
    \emailb
    }
}

\author{Simone Di Giorgio
          \affiliation{
        Istituto di Ingegneria del Mare\\ 
        Consiglio Nazionale delle Ricerche\\
        Via di Vallerano 139, Rome, Lazio, 00128, Italy\\
    \emailc
    }
}

\author{Jannike Solsvik
          \affiliation{
        Department of Chemical Engineering\\
	Norwegian University of Science and Technology\\
	Sem Sælands vei 6, Trondheim, Trøndelag, 7491, Norway\\
    \emaild
    }
}

\author{Shahab Mirjalili
  \affiliation{
        FLOW, Department of Engineering Mechanics\\
	KTH Royal Institute of Technology\\
	Osquars backe 18, Stockholm, Stockholms län, 114 28, Sweden\\
    \emaile
    }
}

\begin{document}

\maketitle   %Print title matter
\thispagestyle{fancy}

% Set the font to 9pt.
\fontsize{9}{11}\selectfont
%\captionsetup{font=small,{small, stretch=11}}
\captionsetup[sub]{font=small}

%%%%%%%%%%%%%%%%%%%%%%%%%%%%%%%%%%%%%%%%%%%%%%%%%%%%%%%%%%%%%%%%%%%%%%
\section*{ABSTRACT}
High-fidelity numerical simulations of bubble-laden, vertical channel flow in the upward configuration, where the bubbles undergo topological changes (breakup and coalescence), were performed with the purpose of exploring the effect of the surface tension on the turbulence anisotropy in the carrier phase. A qualitative analysis shows that velocity fluctuations are enhanced in the wake of large bubbles. Moreover, as shown by the velocity spectra, these structures seem to scale with bubble size and interact with those closer to the wall. Finally, a barycentric map and other indicators of turbulence anisotropy clarify that, except at the core of the channel where the largest bubbles reside, the multiphase flow cases are actually more isotropic than the single-phase flow at a matching friction Reynolds number. This unexpected behavior is attributed to a better redistribution of energy due to an enhancement of sweep events (high-speed fluid towards the wall) in the presence of large bubbles.
%%%%%%%%%%%%%%%%%%%%%%%%%%%%%%%%%%%%%%%%%%%%%%%%%%%%%%%%%%%%%%%%%%%%%%
\section*{INTRODUCTION}
Turbulent bubbly flows are found in a multitude of settings. 
{\color{black}{From wave breaking on the surface of oceans to bubble column reactors and piping systems used in industrial applications.
In many instances}}, these flows present two-way coupled dynamics between the phases, making their analysis quite complex. 
In particular, in the case of bubbly flow delimited by walls, velocity fluctuations in the carrier may arise not only due to the action of an external pressure gradient or imposed flow rate--as in single-phase flow--but also due to buoyancy and the generation of bubble-induced agitation~\citep{r1}. In addition, other effects such as interface contamination, gas volume fraction, bubble deformability, and topological changes in the dispersed phase (i.e., breakage and coalescence) may affect the bubbly flow motion. 

Regarding the study of turbulent bubbly flow that undergoes topological changes, there is in fact a gap in the literature. Many experiments and nearly all numerical simulations have focused on dispersed flow where {\color{black}bubbles, for the most part, do not go through changes in topology}~\citep{r2}. It was not until fairly recently that works on the statistical description of {\color{black}turbulent} bubbly flows with bubbles undergoing breakup and coalescence started to be reported~\citep{r3, r4, r5}. Furthermore, the focus of these studies has been the investigation of low-order statistics (e.g., mean values, variances, and covariance) with little attention to proper characterization of the states of turbulence{\color{black}; information} that is key for the validation and advancement of second-moment closures for the modeling of turbulent bubbly flow~\citep{r6}. {\color{black}In this context,~\cite{r16} recently conducted the first investigation on turbulence anisotropy in heterogeneous bubbly flow with topological changes. Here, we expand the qualitative discussion, including velocity spectra, and provide a summary of the key findings.} %It is worth commenting that, to properly compare with single-phase turbulence, we focus on the states of turbulence in the carrier-phase only.

\section*{NUMERICAL SIMULATIONS}
\begin{figure}%[ht!]
\centering{
\includegraphics[width=7.5 cm,keepaspectratio]{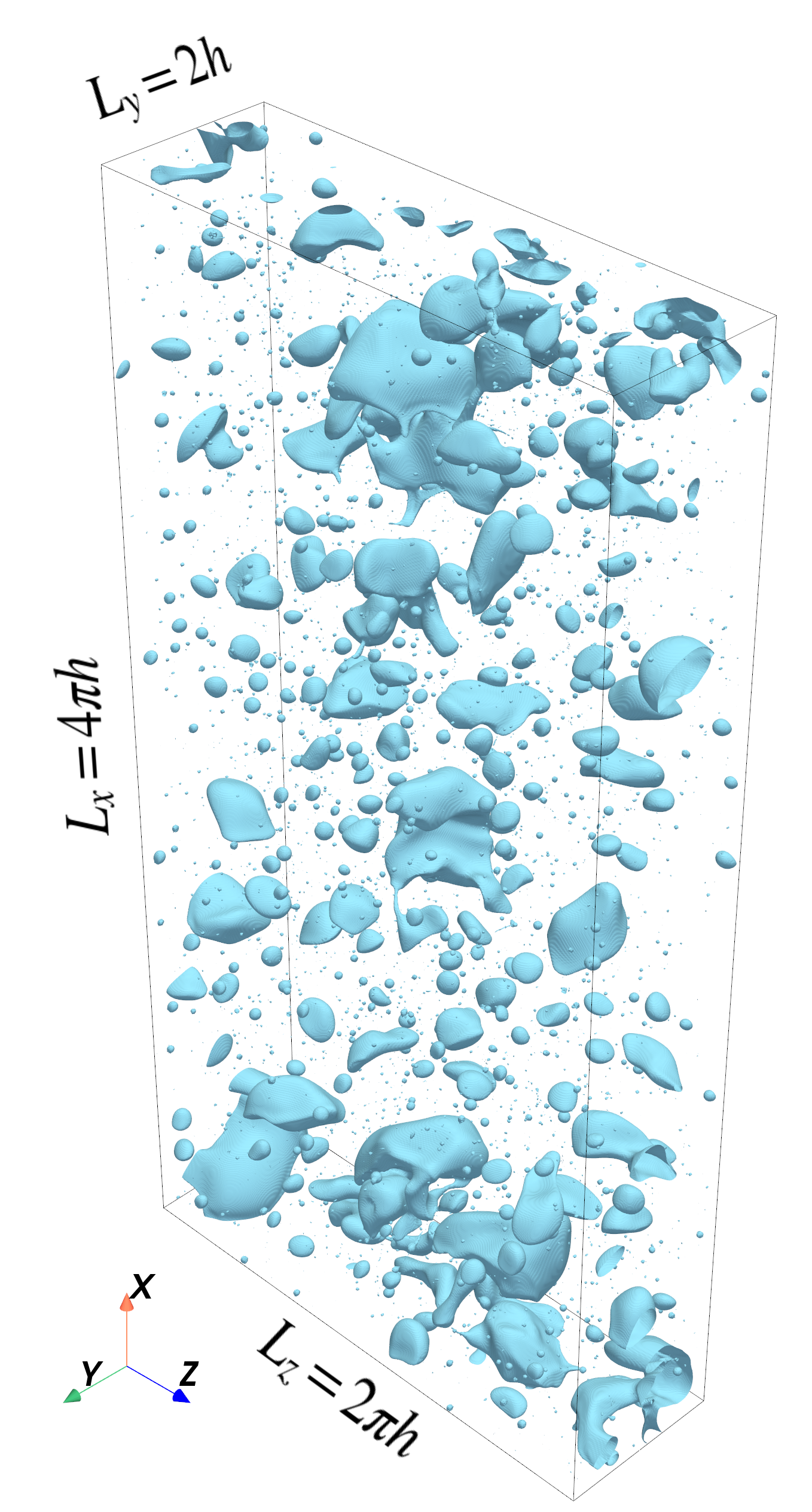}}
\caption{{\fontsize{9}{11}\selectfont Computational domain and instantaneous rendering of the bubbles at $t^+ \approx 10^3$, after discarding the initial transients, for $We_\tau =0.7$.}}
\label{f1}
\end{figure}

{\color{black}High-fidelity numerical simulations of a three-dimensional, bubble-laden vertical channel in the up-flow configuration were performed}. See figure~\ref{f1}. The computation domain has a size of $4\pi h \times 2 h \times 2\pi h$ in the $x$ (streamwise), $y$ (wall-normal) and $z$ (spanwise) directions, respectively. Note that, $h$ (half-channel width) is the characteristic length and $u_\tau$ (friction velocity) is the characteristic velocity scale. The flow is driven by an effective constant pressure gradient that also accounts for the weight of the mixture. Periodicity is imposed for all variables in $x$ and $z$, while {\color{black} the no-slip and impermeability boundary conditions are imposed for the velocity field at the walls}. A right constant angle ($\pi/2$) is imposed for the phase indicator. {\color{black}The code} solves the one-fluid incompressible Navier--Stokes equations and the interface is captured using a volume-of-fluid (VOF) method~\citep{r7}. Details about the numerical implementation and a thorough validation of the code are found in~\cite{r8} and~\cite{r9}, respectively.  

In this problem, the important physical parameters are $Re_\tau$ (friction Reynolds number; ratio of inertial to viscous forces), $We_\tau$ (friction Weber number; ratio of inertial to surface tension forces) and $Fr$ (Froude number; ratio of inertial to buoyancy forces). In addition{\color{black}, the multiphase flow is also characterized by} $\alpha$ (gas volume fraction), $\rho_r$ (density ratio; bubbles to carrier phase), and $\mu_r$ (dynamic viscosity ratio; bubbles to carrier phase). With the objective of studying turbulence anisotropy with surface tension--thus, for cases with differences in the breakage and coalescence events--multiphase simulations with three different $We_\tau$ were carried out: $2.8$, $1.4$, and $0.7$. All other parameters were kept constant: $Re_\tau =150$, $Fr = 0.014$, $\alpha = 5.4\%$, $\rho_r = 0.1$, and $\mu_r =1$. The results of a single-phase (SP) simulation at $Re_\tau = 150$ serve as a basis for comparison with the multiphase cases.

It is also noted that a fixed grid resolution was used in all simulations. For the two homogeneous directions, a constant grid spacing of $\Delta x^+=\Delta z^+=1.84$ was considered. On the other hand, in the wall-normal direction, a natural grid stretching was employed~\citep{r10}, with a minimum resolution near the wall of $\Delta y^+_{min}=0.041$ and a maximum resolution at the channel center of $\Delta y^+_{max}=1.76$. Here, the superscript $+$ is used to denote viscous units. In the following, to facilitate comparison with SP and homogeneous bubbly channel flow, we only show information corresponding to the largest and smallest $We_\tau$ cases, that is, $2.8$ and $0.7$. For a more complete report of the different cases and for further details about the simulations performed, see~\cite{r16}.  
%%%%%%%%%%%%%%%%%%%%%%%%%%%%%%%%%%%%%%%%%%%%%%%%%%%%%%%%%%%%%%%%%%%%%%
\section*{RESULTS}
Before discussing the states of turbulence in the carrier, we consider some relevant qualitative features of the multiphase flow.  

\subsection*{Qualitative features of the bubbly flow}
\begin{figure}%[ht!]
    \centering
    \begin{subfigure}{0.5\textwidth}
        \caption{}
        \includegraphics[width=7.5 cm,keepaspectratio]{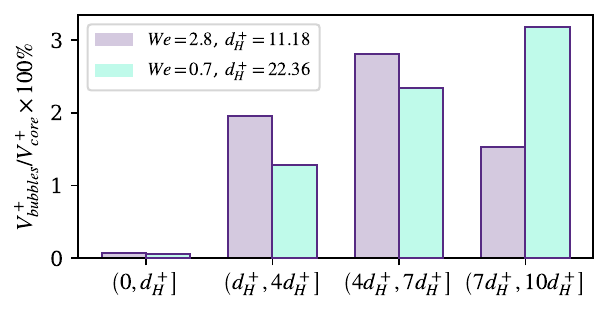}
        
    \end{subfigure}
    
%    \vspace{1cm} % Optional vertical space
    
    \begin{subfigure}{0.5\textwidth}
        \caption{}
        \includegraphics[width=7.5 cm,keepaspectratio]{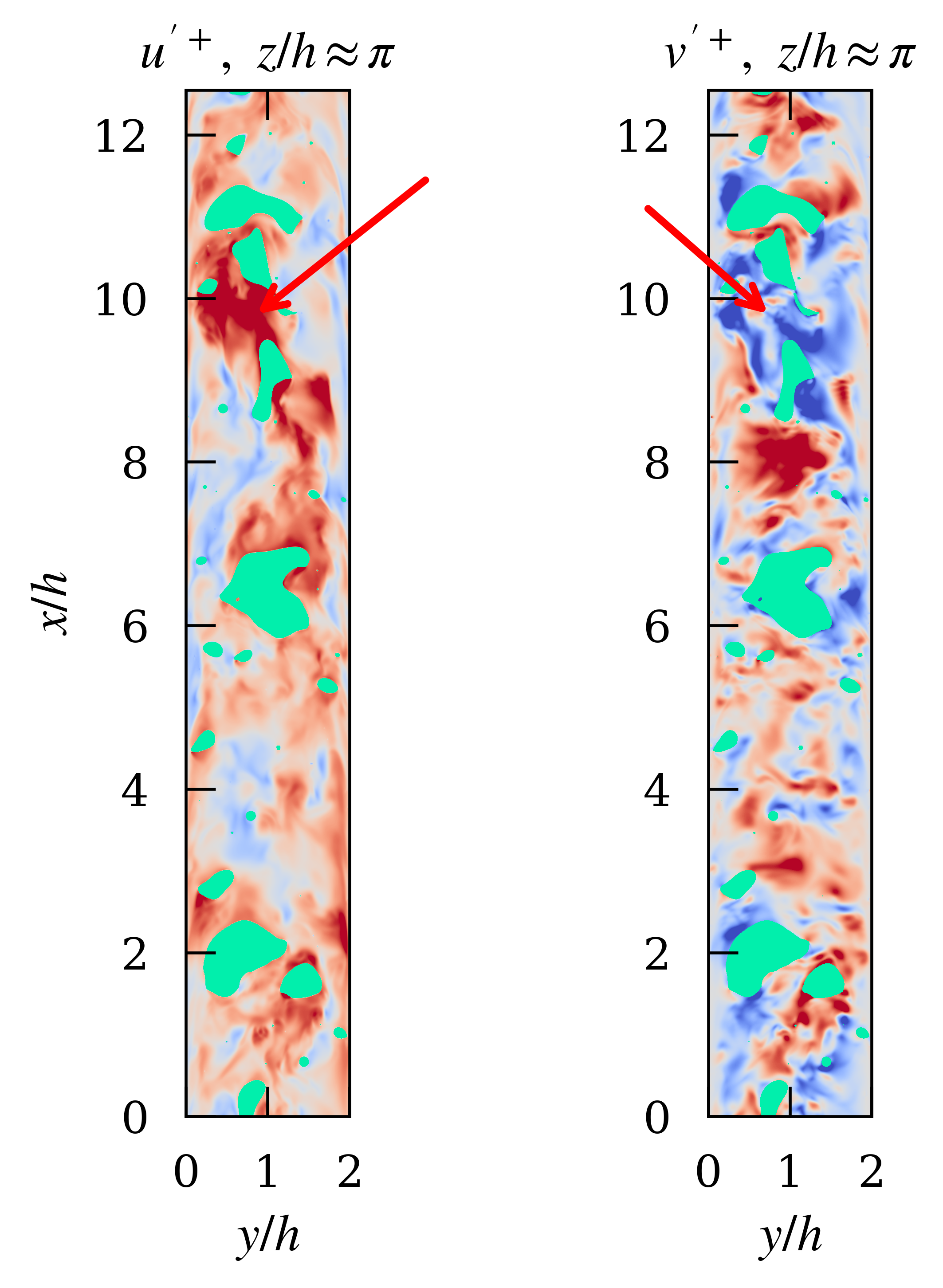}
    \end{subfigure}
    \caption{{\fontsize{9}{11}\selectfont (a) Comparative bar chart for gas volume fraction in the channel core for the different bubbles sizes. (b) Isocontours for streamwise (left) and wall-normal (right) velocity fluctuations in the channel center for $We_\tau = 0.7$. Results are for $t^+ \approx 10^3$, after discarding the initial transients. In (a), $V_{core} = 16\pi^2 h^3/3$. In (b), positive and negative isocontours are respectively identify in red and blue,  while bubbles are rendered in green. Also, the red arrows mark some regions of intense fluctuations in the wake of bubbles.}}
    \label{f2}
\end{figure}

First, as displayed in figure~\ref{f1}, most of the gas volume fraction appears to be encountered in the channel core. This is expected for heterogeneous bubbly flow, as large, deformable bubbles tend to cluster at the center of the channel~\citep{r11}. This also suggests important bubble-induced agitation at the core of the channel. Moreover, such an effect is probably more pronounced with increasing surface tension (decreasing $We_\tau$) due to the appearance of even larger bubbles. {\color{black}Indeed, as shown by the comparative bar chart in figure~\ref{f2}{(a)}, the highest values of $V_{bubble}^+$ are found within the larger bubble sizes; that is, those having $d_{eq}^+$ (bubble equivalent diameter) larger than $d_H^+$~\citep[Hinze scale, the smallest dimension for a stable fluid particle;][]{r12}, and for the highest surface tension case. Here, $V_{bubble}^+$ is the total bubble volume within a range of bubble sizes in the core of the channel; $100\leq y^+\leq 200$.} Furthermore, as shown in figure~\ref{f2}{(b)} for $We_\tau = 0.7$ at the channel center, both ${u^\prime}^+$ (streamwise velocity fluctuations) and ${v^\prime}^+$ (wall-normal velocity fluctuations) are enhanced in the wake of large bubbles. The former mostly in the streamwise direction, whereas the latter both inwards and outwards with respect to the wall-normal direction. It is also noted that the strong, instantaneous velocity structures seem to roughly scale with the bubble size.   

\begin{figure}[ht!]
    \centering
    \begin{subfigure}{0.5\textwidth}
        \caption{}
        \includegraphics[width=7.5 cm,keepaspectratio]{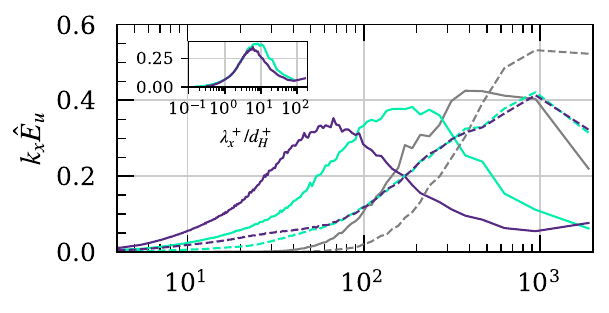}
        
    \end{subfigure}
    
%    \vspace{1cm} % Optional vertical space
    
    \begin{subfigure}{0.5\textwidth}
        \caption{}
        \includegraphics[width=7.5 cm,keepaspectratio]{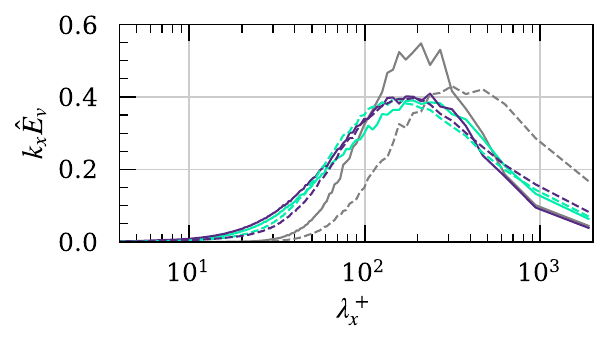}
    \end{subfigure}
    
    \begin{subfigure}{0.5\textwidth}
        \caption{}
        \includegraphics[width=7.5 cm,keepaspectratio]{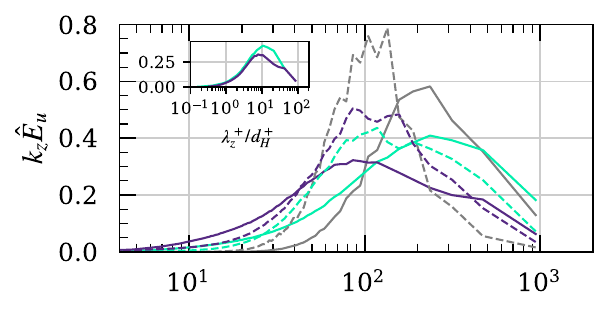}
    \end{subfigure}

    \begin{subfigure}{0.5\textwidth}
        \caption{}
        \includegraphics[width=7.5 cm,keepaspectratio]{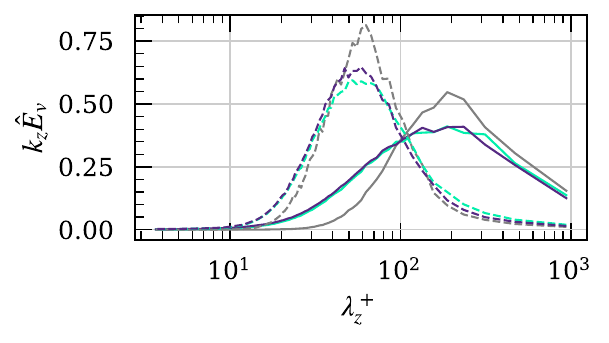}
    \end{subfigure}
    
     \caption{{\fontsize{9}{11}\selectfont Pre-multiplied, normalized, spectral densities for (a), (c) the streamwise velocity component, $\hat E_u$, and (b), (d) the wall-normal velocity components, $\hat E_v$. A gray line style is used for SP, whereas cyan and purple line styles are used for the $We_\tau = 0.7$ and $2.8$ cases, respectively. Continuous- and dashed-line styles respectively identify profiles at $y^+ = 150$ and $y^+ \approx 12$. Note that, $\lambda_x = 2\pi/k_x$ and $\lambda_z = 2\pi/k_z$ are the wavelengths along the homogeneous directions.}}
    \label{f3}
\end{figure}

Next, we consider the streamwise and wall-normal velocity spectra at the center of the channel, as well as at $y^+ \approx 12$, closer to the wall. Figure~\ref{f3} shows the pre-multiplied streamwise and spanwise spectral densities for ${u^\prime}^+$ and ${v^\prime}^+$, which have been normalized by their respective variances, against the corresponding wavelength. As seen from the figure, there is an overall shift of the spectral content towards smaller scales for the bubble-laden cases when compared to SP. This is particularly conspicuous at the channel center, yet it is also seen to a lesser extent closer to the wall. Interestingly, as shown in panels (a) and (c), there is indeed a correspondence between the scales of the most energetic streamwise velocity structures and the bubble size at the core of the channel. In fact, overlapping at low to moderate wavelengths is recovered for both $We_\tau$ cases when normalizing by the respective ${d_H}^+$. See insets. On the other hand, as displayed in panels (b) and (d), the effect of increasing the surface tension (decreasing $We_\tau$) and having larger bubbles on the wall-normal velocity spectra seems negligible. Moreover, as shown in panel (b), very similar streamwise spectral content distributions are seen near the wall and at the channel center, hinting of potentially important interactions between the near-wall structures and those found at the channel core. %The overall shift towards smaller scales at $y^+ \approx 12$ is also tentatively attributed to a near-wall ``footprint'' of the structures found at the channel centre. 

\subsection*{States of turbulence in the carrier phase}
\begin{figure}%[ht!]
    \centering
    \begin{subfigure}{0.5\textwidth}
        \caption{}
        \includegraphics[width=7.5 cm,keepaspectratio]{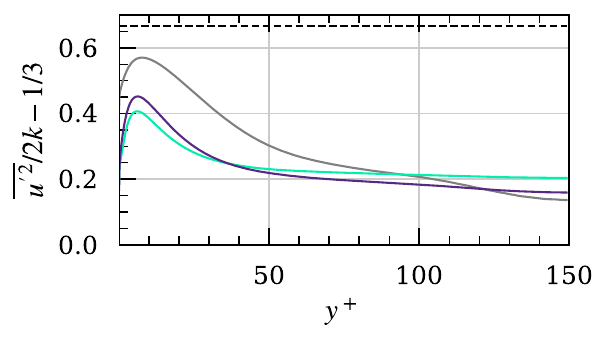}
        
    \end{subfigure}
    
%    \vspace{1cm} % Optional vertical space
    
    \begin{subfigure}{0.5\textwidth}
        \caption{}
        \includegraphics[width=7.5 cm,keepaspectratio]{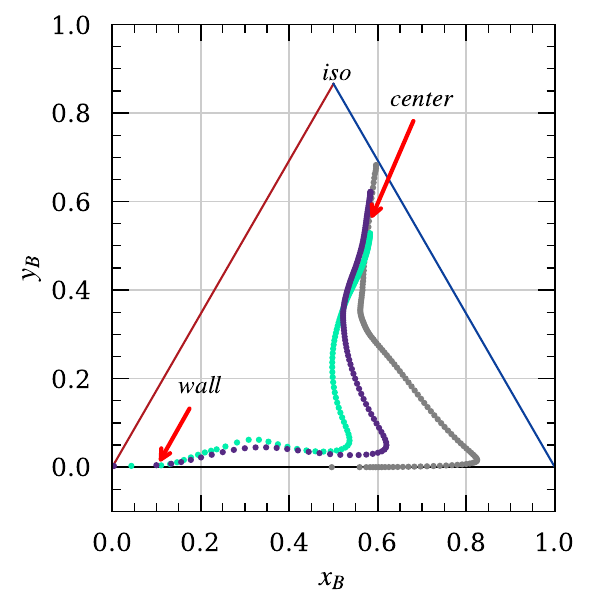}
    \end{subfigure}

    \begin{subfigure}{0.5\textwidth}
        \caption{}
        \includegraphics[width=7.5 cm,keepaspectratio]{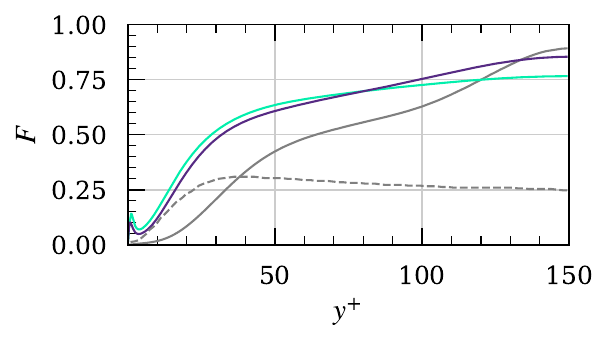}
    \end{subfigure}
    
    \caption{{\fontsize{9}{11}\selectfont (a) First diagonal component of the Reynolds stress anisotropy tensor, $a_{11}$, against $y^+$. (b) Reynolds stress anisotropy in a barycentric map with coordinates $x_B, y_B$. (c) Lumley's flatness, $F$, against $y^+$. {\color{black}Statistics are conditioned} to the carrier phase only. A gray line style/marker is used for SP, whereas cyan and purple line styles/markers are used for the $We_\tau = 0.7$ and $2.8$ cases, respectively. In (a), a dashed-line style identifies the $2/3$ limit. In (b), red, blue, and black line styles are used to identify the axisymmetric expansion, axisymmetric contraction, and two-component limits, respectively. Also, the isotropic limiting state is labeled as $iso$, while the red arrows (with self-explanatory labels) mark states close to the wall and to the channel center. In (c), a dashed-line style identify results corresponding to homogeneous bubbly flow~\citep[case labeled as SmMany in][]{r6}.}}
    \label{f4}
\end{figure}

In order to characterize the states of turbulence in the carrier, we consider the Reynolds stress anisotropy tensor, defined as
\begin{equation}
    a_{ij} = \frac{\overline{u}_{ij}}{2k} -\frac{1}{3}\delta_{ij},
    \label{eq1}
\end{equation}
where $\overline{u}_{ij}$, $k$, and $\delta_{ij}$ are the Reynolds stress tensor, turbulent kinetic energy, and Kronecker delta ($1$ if $i=j$ and $0$ otherwise), respectively. Typically, characterization of turbulent states is performed considering tools such as anisotropy invariant maps, which are based on the invariants of $a_{ij}$~\citep{r17}. Nevertheless, we can also make a rapid diagnosis by directly considering the components of $a_{ij}$. For instance, it is well known that in canonical wall-turbulence, very near the wall ($y^+ \approx 10$) most of the turbulent kinetic energy is found in the streamwise velocity component, whereas near the channel center states closer to isotropic conditions are attained~\citep{r18}. Therefore, $a_{11}$ is expected to approach $2/3$ near the wall and $0$ at the core of the channel. Figure~\ref{f4}(a) shows that this is indeed so for SP. {\color{black}For the bubble-laden cases, states closer and farther away from isotropic conditions are seen at $y^+\approx 10$ and $y^+\approx 150$, respectively}. Furthermore, with the exception of the central channel region, the bubble-laden cases are actually closer to isotropic conditions with respect to SP. 

The previous observation is further corroborated by means of a barycentric map~\citep[BAM;][]{r15} and $F$~\citep[the Lumley's flatness;][]{r13}. See figure~\ref{f4}(b) and (c). A BAM is a linear map in the form of an isosceles triangle, delimitating all realizable states of turbulence. The base represents two-component states, that is, when one normal Reynolds stress component is negligible compared to the others, while the right and left hand sides, respectively, identify the axisymmetric expansion (prolate-like) and axisymmetric contraction (oblate-like) states, which converge at the isotropic ($iso$) limit. For further details on the terminology used to identify the states of turbulence, see, e.g.,~\cite{r19}. On the other hand, the $F$ parameter is an indicator of proximity to the base of the triangle and to the $iso$ limit, taking a value of $1$ and $0$ for isotropic and two-component turbulence, respectively. Then, as shown in panels (b) and (c), turbulence is indeed more anisotropic for the bubble-laden cases at the channel center. This is expected considering the previous observation about larger bubbles residing in the core of the channel and the consequent appearance of strong bubble-induced fluctuations. Rather surprising, on the other hand, is the fact that this trend is truly only confined to the channel center, and as we move from it, there is actually a higher degree of turbulence anisotropy for SP than for heterogeneous bubbly flow. See profiles that are mostly to the left and above the SP one in panels (b) and (c), respectively. It is also interesting to note that this behavior--higher degree of isotropy when compared to SP--is also quite different from the one reported for homogeneous bubbly channel flow. See representative case corresponding to a monodisperse swarm of small bubbles~\citep[case labeled as SmMany in][]{r6} in panel (c), which displays a very anisotropic and approximately constant state of turbulence across a significant portion of the channel.

\subsection*{Sweeps and ejections}
\begin{figure*}%[ht!]
    \centering
    \includegraphics[width=10 cm,keepaspectratio]{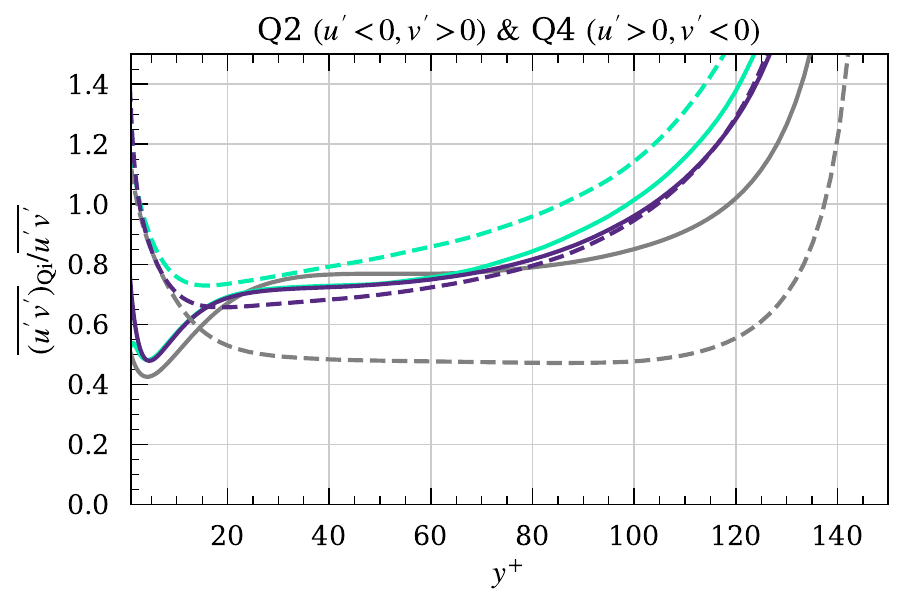}
    \caption{{\fontsize{9}{11}\selectfont Fractional contribution of {\color{black}ejections (Q2) and sweeps (Q4)} to the shear Reynolds stress, $\overline{(u^{\prime}v^{\prime})}_\text{Qi}/\overline{u^\prime v^\prime}$, against $y^+$. A gray line style is used for SP, whereas cyan and purple line styles are used for the $We_\tau = 0.7$ and $2.8$ cases, respectively. Continuous- and dashed-line styles respectively identify the fractional contributions of the {\color{black}Q2 and Q4} events. {\color{black}Statistics are conditioned} to the carrier phase only.}}
    \label{f5}
 \end{figure*}

Lastly, in light of potentially important inner/outer interactions discussed in the qualitative section and with the purpose of providing a plausible physical explanation for the unexpected changes in turbulence anisotropy for the bubbly-laden cases, we analyze intense momentum transfer events in the wall-bounded flow by means of a quadrant analysis~\citep{r20}. Figure~\ref{f5} presents the fractional contribution of sweeps ({\color{black}Q4}, high-speed fluid inward) and ejections ({\color{black}Q2}, low-speed fluid outward) events to $\overline{u^\prime v^\prime}$ (the Reynolds shear stress). These are the quadrant events that positively contribute to $\overline{u^\prime v^\prime}$, while the others (Q1 and Q3, not shown), simply counterbalance the Q2s and Q4s such that at any $y^+$, the fractional contribution of the four quadrant events is unitary. Therefore, as seen in the figure, both sweeps and ejections increase for the bubble-laden cases. However, those that significantly deviate from the SP results for most of the channel are the sweeps. It is also noted that there are some differences between the two bubble-laden cases. For $We_\tau = 2.8$, there is still a crossing point where Q2s overcome Q4 events, whereas this is not the case for $We_\tau = 0.7$. In fact, for the former, the sweeps appear to be as important as the ejections for a significant portion of the channel, while for the latter, the sweeps are always the dominant momentum transfer events. In other words, higher momentum in the form of high-speed fluids towards the wall is present for the bubble-laden cases, which may lead to a better redistribution of energy across the channel and explain the observed decrease in anisotropy seen for most of the channel, when compared to SP. In addition, this effect is more pronounced with increasing surface tension (decreasing $We_\tau$), {\color{black}as larger bubbles appear in the core of the channel and there is more bubble-induced agitation; see the qualitative discussion.} 

%%%%%%%%%%%%%%%%%%%%%%%%%%%%%%%%%%%%%%%%%%%%%%%%%%%%%%%%%%%%%%%%%%%%%%
\section*{CONCLUSIONS}
{\color{black}High-fidelity simulations of bubble-laden vertical channel flow were performed}. The flow is driven upward by an effective constant pressure gradient, and the deformable bubbles undergo breakage and coalescence. The interface is captured by a VOF method. Three sets of simulations, only differing in their $We_\tau$, {\color{black}were} performed to observe the effect of changing the surface tension on the turbulence anisotropy in the carrier. {\color{black}Here, only the results corresponding to $We_\tau = 0.7$ and $2.8$ are shown. As a basis for comparison, a SP case with matching $Re_\tau$ {\color{black}was also considered}. 

{\color{black}The main finding is that the turbulence anisotropy in the carrier is actually lower for the multiphase flow cases compared to SP}. Only close to the core of the channel where the largest bubbles reside is the opposite true. This is also in striking contrast to results reported for homogeneous bubbly channel flow, where the turbulence anisotropy is actually higher for most of the channel compared to SP~\citep{r6}. Such behavior is attributed to an important increase of the momentum transfer events consisting of high-speed fluid towards the wall, which may help to a better redistribution of energy among the different normal Reynolds stress components and across the channel for the considered bubble-laden cases. 

Further analyzes{\color{black}, with the aim of clarifying the inner/outer interactions and the transport of turbulent kinetic energy and momentum,} may include a full reporting of the velocity spectra and the Reynolds stress budgets.

\section*{ACKNOWLEDGMENTS}
Computing resources awarded by EuroHPC (project ID: R0387) and Sigma2 (project ID: NN9646K) are acknowledged. AA and JS also acknowledge financial support by the Research Council of Norway (RCN, grant number: 334652).
%%%%%%%%%%%%%%%%%%%%%%%%%%%%%%%%%%%%%%%%%%%%%%%%%%%%%%%%%%%%%%%%%%%%%%
\bibliographystyle{tsfp}
\bibliography{tsfp}

%%%%%%%%%%%%%%%%%%%%%%%%%%%%%%%%%%%%%%%%%%%%%%%%%%%%%%%%%%%%%%%%%%%%%%

\end{document}